\documentclass[usegraphicx]{mn2e}

\title[A wavelet code for $N$-body simulations and data de-noising]
      {A wavelet add-on code for\\
       new-generation {\boldmath $N$}-body simulations and data de-noising\\
       ({\Large\bfseries JOFILUREN})}

\author[A. B. Romeo, C. Horellou and J. Bergh]
       {Alessandro B. Romeo,$^{1}$\thanks{E-mail: romeo@oso.chalmers.se}
        Cathy Horellou$^{1}$ and
        J\"{o}ran Bergh$^{2}$\\
        $^{1}$Department of Astronomy and Astrophysics,
              Centre for Astrophysics and Space Science,
              Chalmers University of Technology,
              \\
              SE-43992 Onsala, Sweden\\
        $^{2}$Department of Mathematics,
              Chalmers University of Technology and G\"{o}teborg University,
              SE-41296 G\"{o}teborg, Sweden}

\begin{document}

\date{Accepted ......... .  Received ......... ; in original form .........}

\pagerange{\pageref{firstpage}--\pageref{lastpage}}

\pubyear{2004}

\maketitle

\label{firstpage}

\begin{abstract}
Wavelets are a new and powerful mathematical tool, whose most celebrated
applications are data compression and de-noising.  In Paper I (Romeo,
Horellou \& Bergh 2003), we have shown that wavelets can be used for removing
noise efficiently from cosmological, galaxy and plasma $N$-body simulations.
The expected two-orders-of-magnitude higher performance means, in terms of
the well-known Moore's law, an advance of more than one decade in the future.
In this paper, we describe a wavelet add-on code designed for such an
application.  Our code can be included in common grid-based $N$-body codes,
is written in Fortran, is portable and available on request from the first
author.  The code can also be applied for removing noise from standard data,
such as signals and images.
\end{abstract}


\section{INTRODUCTION}

In $N$-body simulations the number of particles $N$ cannot generally be set
equal to the number of bodies of the real system, but is dictated by the
available computer power.  A simulation with, say, ten times more particles
demands at least one order of magnitude more computational time, memory and
storage.  Because of such a limitation, $N$ is generally several orders of
magnitude smaller than required.  A small $N$ means that the statistical
fluctuations of particle positions and velocities are artificially enhanced,
and so are collisional effects.  This is dangerous because collisions affect
the ability of resonances to damp or amplify perturbations, which in turn
affects the formation of structures and the dynamical evolution of the model.
Thus an effective method of noise reduction is required.

   The standard way to reduce noise in $N$-body simulations is to soften the
interparticle force at short distances, either directly or using finite-sized
particles (e.g., Romeo 1994, 1997, 1998a,\,b; Dehnen 2001; see also Byrd
1995).  On the other hand, softening reduces noise only in part.  The initial
conditions imposed on particle positions and velocities are also relevant.
Basically, noise can be suppressed at the beginning of the simulation by
sampling phase space regularly (quiet starts), rather than randomly (noisy
starts), consistent with the distribution function of the model (e.g., Dawson
1983; Birdsall \& Langdon 1991; Knebe, Green \& Binney 2001).  Even so, noise
will develop during the simulation.  In fact, quiet starts impose an initial
order on the model.  But the model will react to such a state of low entropy
and follow the natural tendency of physical systems towards thermalization.
The development of noise is mediated by instabilities, which amplify and
randomize the initial correlations arising from the discrete regular sampling
of phase space.

   The effects of noise are subtle and not yet fully understood.  Today, half
a century after the first $N$-body simulations, there is still an intense
debate.  Noise is a crucial issue for simulations of structure formation in
the early Universe (e.g., Splinter et al.\ 1998; Hamana, Yoshida \& Suto
2002; Power et al.\ 2003; Binney 2004; Diemand et al.\ 2004; Sylos Labini,
Baertschiger \& Joyce 2004), and for galaxy simulations (e.g., Pfenniger
1993; Pfenniger \& Friedli 1993; Weinberg \& Katz 2002; O'Neill \& Dubinski
2003; Valenzuela \& Klypin 2003).  Noise is an important issue not only for
cosmology and astrophysics but also for plasma and accelerator physics, where
simulations are used for technological applications such as fusion and
charged particle beams (e.g., Dawson 1983; Birdsall \& Langdon 1991; Arter
1995; Kandrup 2003).  \emph{The noise problem is acute and awaits solution.}

   Wavelets are the state-of-the-art technique used for noise reduction in
digital signal/image processing (see, e.g., Mallat 1998; Bergh, Ekstedt \&
Lindberg 1999; for traditional techniques such as data averaging or Wiener
filtering see, e.g., Gonzalez \& Woods 2002).  Wavelets have an intrinsic
ability to compress the signal into few large coefficients, so that noise can
be removed with a proper thresholding.  Being intrinsic, their ability is
independent of general properties of the data such as the number of
dimensions or the presence of symmetries.  Wavelet de-noising is very
effective: it outperforms traditional techniques of noise reduction and the
algorithm is even faster than the fast Fourier transform.  The second aspect
is especially important for our application since speed is a primary factor
in $N$-body simulations.  We recommend the following literature: for the
continuous and fast wavelet transforms see Addison (2002) and Goedecker
(1998), respectively; for physical applications see again Goedecker (1998)
and the beautiful book by van den Berg (2004).

   In Paper I, we have pioneered the first application of wavelet de-noising
to $N$-body simulations.  Our method has been subjected to several hard
tests.  The conclusion is that it can make the simulation equivalent to a
simulation with two orders of magnitude more particles.  The implications are
clear.

   In the present paper, we show that our method even allows controlling the
effectiveness of de-noising: the simulation can be made equivalent to a
simulation with $\Gamma$ times more particles, where $\Gamma$ is assigned by
the simulator beforehand.  Such a degree of freedom can be exploited for
understanding the effects of noise more thoroughly.  Besides, we describe the
code that implements our method.  It is an add-on code, and as such is meant
to be included in the simulator's $N$-body code.  This is simple if the
$N$-body code is of particle-mesh type.  Our code can also be used by itself
for de-noising standard data, such as signals and images.  It is written in
Fortran, is portable and available on request from the first author.  Last
but not least, we have the ambition to provide a reader-friendly and
self-contained discussion of wavelet de-noising, from the basics to the most
advanced aspects of our problem.  For further reading see Paper I and the
literature already recommended.

   The rest of our paper is organized as follows.  Wavelets and wavelet
applications are overviewed in Sects 2 and 3, respectively.  Sect.\ 3.2 not
only discusses the basics of de-noising, but also explains the steps of the
de-noising algorithm in our code.  These are then discussed in detail in
Sect.\ 4.  After such a tour of the code, we explore advanced de-noising in
Sect.\ 5.  There we learn how to control the effectiveness of de-noising, and
how this part of the method is implemented in the code; besides, we discuss
further aspects of the problem.  Practical points concerning the use of the
code are discussed in Sect.\ 6.  Finally, the conclusions are drawn in Sect.\
7.

\section{BASICS OF WAVELETS}

\subsection{The fundamental property of wavelets}

Data such as signals, images and those arising from the numerical solution to
physical problems generally enclose information on various scales.  In order
to extract such information, we should be able to separate small-scale
features from large-scale features and to understand their contributions to
the overall structure of the data.  The classical technique used for this
purpose is the Fourier transform, which encodes the original time/space
information into frequency content of the data, the frequency being roughly
the inverse of the relevant scale.  But the Fourier transform runs into a
serious difficulty: it loses all information about the time/space
localization of a given frequency component.  This is nothing but the
consequence of the Heisenberg uncertainty principle in the context of data
processing.

   The traditional way to overcome this difficulty is to localize the complex
sinusoid of the transform multiplying it by a window function, a Gaussian for
example, which is then translated across the data.  For a window of given
shape, its width determines not only the time/space resolution but also the
frequency resolution, again as a consequence of the Heisenberg uncertainty
principle.  A narrow window gives a good time/space resolution but a bad
frequency resolution, and vice versa for a wide window.  So how should we
choose the width of the window?  If we choose it comparable to the smallest
scale of interest, the time/space resolution of course matches the data, but
the frequency localization is too poor to resolve the low frequencies
characterizing large-scale features.  And if we choose a wider window so as
to have a finer frequency resolution, the time/space resolution gets too
coarse to analyse small-scale features.  Thus even the windowed Fourier
transform runs into a difficulty: it has a fixed time/space-frequency
resolution [constant bandwidth, in the language of data processing; cf.\
Paper I, fig.\ 1 (left-hand panel)].

   Wavelets are a multi-scale method that overcomes this difficulty.  Their
fundamental property is to provide an \emph{adaptive} time/space-frequency
resolution, in the sense that the uncertainty in frequency is proportional to
the frequency itself [constant relative bandwidth, in the language of data
processing; cf.\ Paper I, fig.\ 1 (right-hand panel)].  In other words, this
means that small-scale features of the data are analysed with fine resolution
in time/space and coarse resolution in frequency, as is natural, and vice
versa for large-scale features.

\begin{figure*}
\scalebox{1.}{\includegraphics{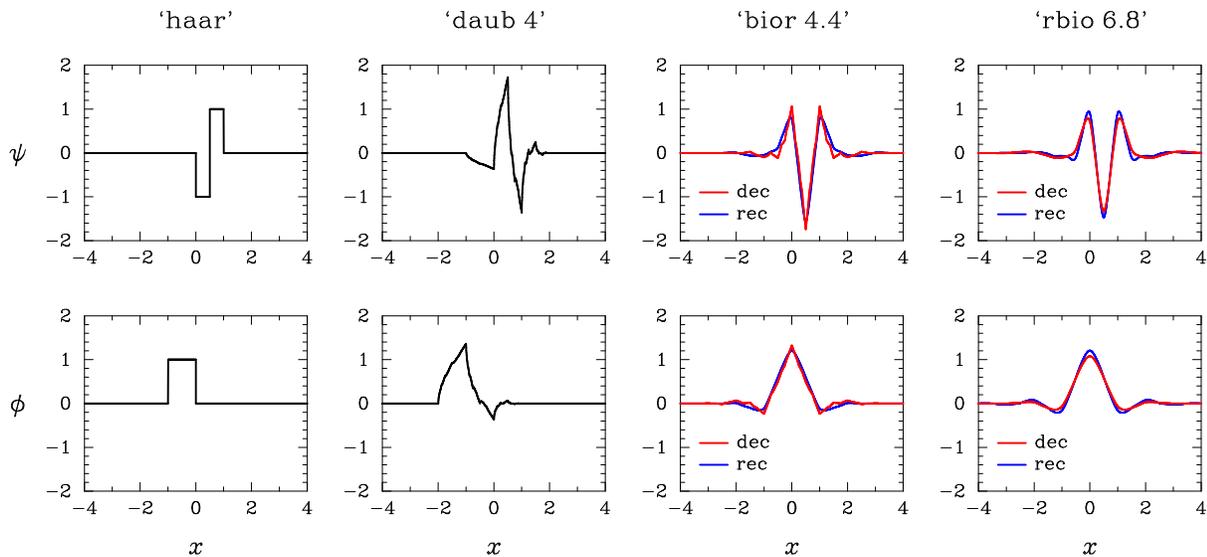}}
\caption{Various wavelets $\psi(x)$ and scaling functions $\phi(x)$.  The
         wavelets `haar' and `daub\,4' are orthogonal.  The wavelet pairs
         `bior\,4.4' and `rbio\,6.8' are bi-orthogonal and also
         quasi-orthogonal, hence for each of them the decomposition (`dec')
         and reconstruction (`rec') wavelets are distinct but similar.
         Analogous considerations apply to the scaling functions.}
\end{figure*}

\subsection{Wavelet transform}

In order to provide an adaptive time/space-frequency resolution, the wavelet
analysis involves localized wave-like functions, which are contracted or
dilated over the relevant range of scales and translated across the data.  On
the other hand, there are several ways to carry out the analysis, depending
on whether the data are continuous or discrete; and, in the discrete case,
depending on technical factors.  Here we present the wavelet transform that
is most appropriate for our application, which is also the one most commonly
used for compressing and de-noising discrete data.

   The contributions of small-scale and large-scale features are singled out
with an iterative procedure.  The first step consists of separating the
smallest-scale features from the others.  It is done by passing the data
through a high-pass filter and a complementary low-pass filter.  These
filters are the discrete counterparts of the analysing functions of the
transform, the wavelet $\psi(x)$ and the scaling function $\phi(x)$
respectively, and are constructed with a mathematical technique known as
multi-resolution analysis.  Filtering produces redundant information, since
each set of filtered data has the same size as the original data.  Redundancy
is avoided by rejecting every other point of the filtered data.  It is well
known that down-sampling produces aliasing in the context of the Fourier
transform, but the filters of the wavelet transform are constructed in such a
way as to eliminate it.  The second step consists of separating the features
that appear on a scale twice as large as in the first step.  It is done by
regarding the low-pass filtered and down-sampled data as new input data, and
by analysing them as in the first step.  The procedure continues until the
largest-scale features are also separated.  In summary, the wavelet transform
decomposes the original data into a coarse approximation and a sequence of
finer and finer details, keeping the total size of the data constant (cf.\
Paper I, fig.\ 2).  We can draw an analogy with art and say that the
approximation gives an `Impressionist' view of the data!

   The original data can be reconstructed with the inverse wavelet transform.
The coarsest approximation and detail are up-sampled, filtered and added.
Here up-sampling means inserting zeros between the data points, and the
filters are closely related to the decomposition filters so as to eliminate
aliasing.  The output is a finer approximation, which is then combined with
the corresponding finer detail as above, and the procedure is iterated.  In
practice, the wavelet synthesis is carried out for reconstructing data that
have been processed in wavelet domain, such as in data compression and
de-noising.

   The great success of the wavelet transform arises not only from its
adaptive time/space-frequency resolution, but also from its speed: it is even
faster than the fast Fourier transform.  Given data of size
$N_{\mathrm{d}}=2^{J}$ ($J$ positive integer) and filters of effective size
$2M$, the fast wavelet transform is computed with $4MN_{\mathrm{d}}$
arithmetic operations, where $M$ is a positive integer independent of
$N_{\mathrm{d}}$ and typically smaller than ten.  In contrast, the fast
Fourier transform has complexity $2N_{\mathrm{d}}\log_{2}N_{\mathrm{d}}$.
Note that such an efficiency follows directly from the non-redundancy of the
transform.  Similar considerations apply to the inverse fast wavelet
transform.

\begin{figure*}
\scalebox{.97}{\includegraphics{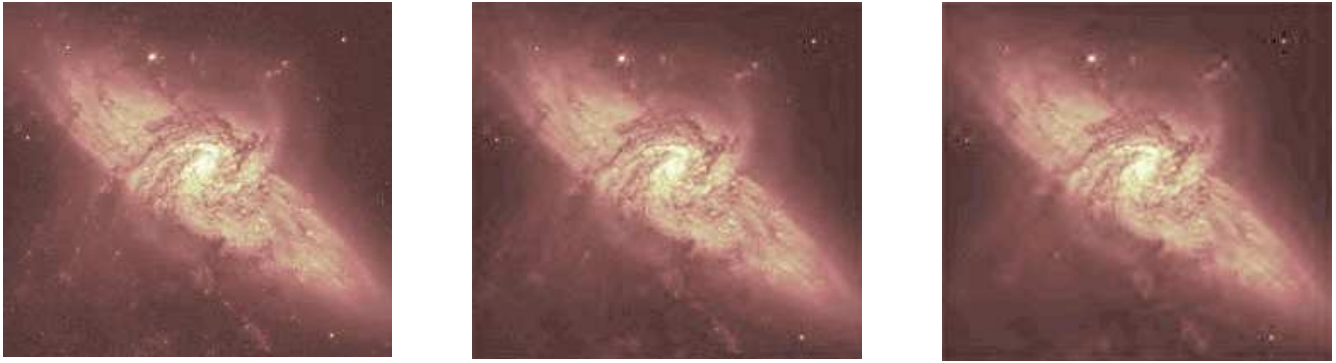}}
\caption{The original image of the spiral galaxy pair NGC 3314 (left) is
         compressed by a factor of 200 with only 0.29\% loss of information
         (middle) and by a factor of 500 with only 0.52\% loss of information
         (right).  The original image is from Hubble Heritage, courtesy of
         NASA and STScI.}
\end{figure*}

\subsection{Wavelet properties}

In contrast to classical transforms, where the analysing functions belong to
a single class and are defined analytically, there are dozens of wavelet
families and their members are generally defined numerically through the
associated filters.  Why do we have so many choices?  Because, even though
the fast wavelet transform has an adaptive time/space-frequency resolution,
there are various ways to optimize the trade-off between time/space and
frequency localizations, and different conditions can be imposed.  In other
words, wavelets are not all equivalent in applications and, if we want to
choose the optimal wavelet for a given problem, we must understand their
properties well.  The properties of the scaling functions are determined by
those of the wavelets, but are themselves less relevant.  The wavelet
properties are: size of support, symmetry, number of vanishing moments,
regularity and (bi-)orthogonality.

\begin{description}
   \item \emph{Size of support.}  The support of a wavelet is the interval
where the wavelet is non-zero.  Its size determines not only the time/space
localization of the wavelet, but also the speed of the transform.
   \item \emph{Symmetry.}  Symmetry also influences the quality of
time/\-space localization.  For example, an asymmetric wavelet can be
regarded as giving a location with asymmetric error bars.
   \item \emph{Number of vanishing moments.}  A wavelet $\psi(x)$ has $n$
vanishing moments when
\begin{equation}
\int_{-\infty}^{+\infty}x^{\nu}\psi(x)\mathrm{d}x=0\mbox{\ \ \ \ for\ \ }\nu=
0,1,\ldots,n-1\,;
\end{equation}
where $x$ denotes time or space.  In particular, all `normal' wavelets have
zero mean ($n=1$) since, under rather general assumptions, this is related to
the admissibility condition for the existence of the inverse transform.  The
number of vanishing moments affects the frequency localization.  In fact, the
Fourier transform of a wavelet with $n$ vanishing moments peaks at a
characteristic frequency and decays as $k^{n}$ towards the origin, where $k$
denotes frequency.
   \item \emph{Regularity.}  Regularity also affects the frequency
localization.  In fact, the Fourier transform of a wavelet that is continuous
together with its first $n-1$ derivatives decays as $k^{-(n+1)}$ towards
infinity.
   \item \emph{(Bi-)Orthogonality.}  The orthogonality property concerns the
set of wavelets defining the transform, that is the set of scaled and
translated versions of the basic wavelet.  It means that such wavelets form
an orthogonal basis.  The alternative bi-orthogonality property means that
the decomposition and reconstruction wavelets form two distinct bases, which
are mutually orthogonal.  Note that (bi-)orthogonality is intimately related
to the non-redundancy of the transform.
\end{description}

   It follows that a good time/space localization requires a small support
and high symmetry, and a good frequency localization requires many vanishing
moments and high regularity.  A small support is also needed for a faster
transform.  On the other hand, the wavelet properties are interrelated.  A
small support implies relatively few vanishing moments and low regularity.
In addition, orthogonality implies asymmetry, except for the simplest
wavelet.  Bi-orthogonality weakens the coupling between the properties of the
decomposition and reconstruction wavelets, and allows perfect symmetry.  This
means that the requirements above cannot be satisfied equally well.  In order
to choose a good wavelet, we should then know their relative importance,
which depends on the application.

\subsection{What do wavelets look like?}

Let us now illustrate what wavelets and scaling functions look like.  (Recall
that the scaling functions are the continuous counterparts of the low-pass
filters of the fast wavelet transform and its inverse; see Sect.\ 2.2, and
also Sect.\ 2.3.)  Fig.\ 1 shows various representatives.  The wavelets
`haar' and `daub\,4' belong to the family of Daubechies wavelets, and are
simple (see Daubechies 1992; here we use the same names as in the code).
These wavelets are orthogonal; and they have few vanishing moments, small
support, low regularity and no symmetry (`haar' is an exception).  The
wavelet pairs `bior\,4.4' and `rbio\,6.8' belong to the family of
bi-orthogonal spline wavelets and to its reverse, respectively, and are more
advanced (see Daubechies 1992; here we use the same names as in the code).
Such wavelet pairs are not only bi-orthogonal but also quasi-orthogonal: for
each pair the decomposition and reconstruction wavelets are distinct but
similar.  In addition, they have more vanishing moments, larger support,
higher regularity and perfect symmetry.

\section{BASICS OF WAVELET APPLICATIONS}

\begin{figure}
\centering
\scalebox{.79}{\includegraphics{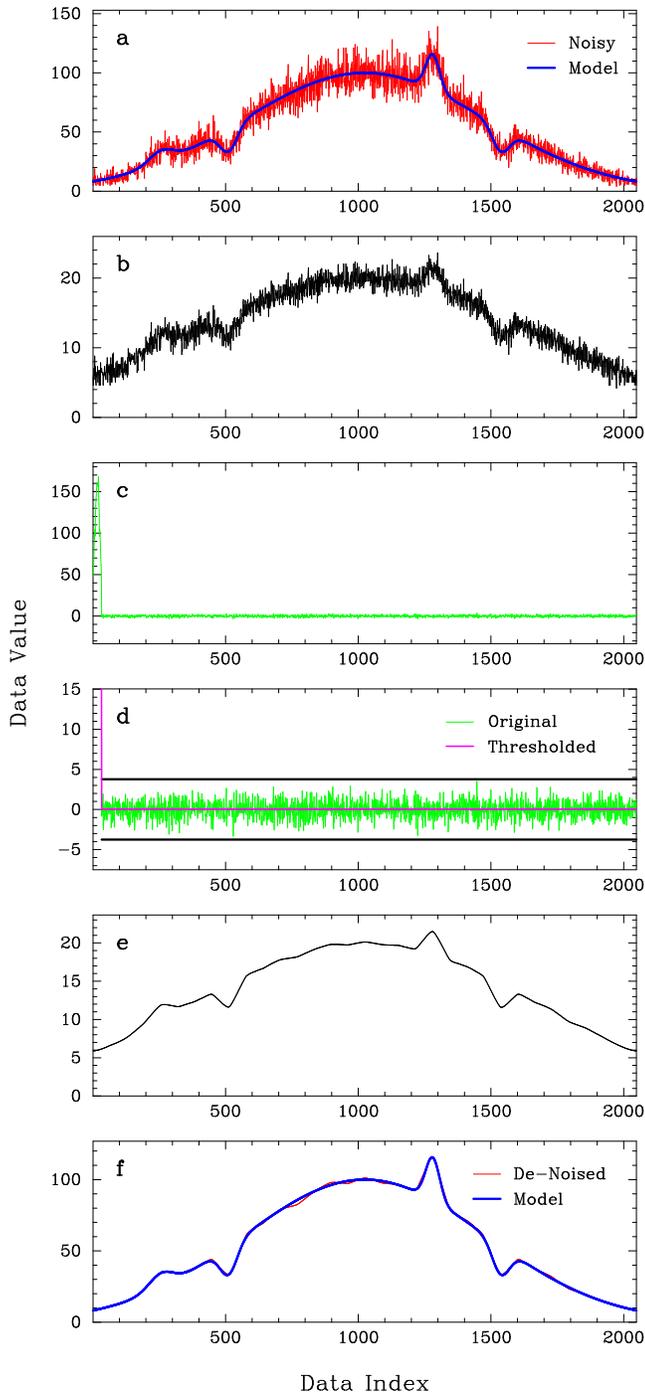}}
\caption{De-noising in action.  \textbf{a}: data with Poissonian noise and
         the perfectly non-noisy model data.  \textbf{b}: pre-processed data.
         \textbf{c}: wavelet coefficients computed by fast wavelet
         transforming, after the choice of the wavelet; note that there are
         few large wavelet coefficients and many small wavelet coefficients.
         \textbf{d}: original and thresholded wavelet coefficients; the
         threshold is also shown.  \textbf{e}: data computed by inverse fast
         wavelet transforming.  \textbf{f}: de-noised data computed by
         post-processing vs.\ the model data.  The signal-to-noise ratio is
         $\mathit{SNR}\simeq8.8$ in the noisy data and
         $\mathit{SNR}\simeq73.9$ in the de-noised data, hence the de-noising
         factor is $\mathit{DF}\simeq8.4$.  (For more information see the
         text.)}
\end{figure}

\subsection{Data compression}

The adaptive time/space-frequency resolution and the non-redundancy of the
fast wavelet transform have an important implication: given regular data,
most information present in them gets concentrated into few large wavelet
coefficients.  In practice, this means that we can set all the other
coefficients to zero and get back data almost identical to the original ones.
This is the idea behind data compression.

   The compression ability can be quantified by the compression factor
$\mathit{CF}$ and the loss of information $\mathit{LI}$, defined as:
\begin{equation}
\mathit{CF}=\frac{N_{W}}{n_{w}}\,,
\end{equation}
\begin{equation}
\mathit{LI}\,[\%]=100\left(1-\frac{{\displaystyle\sum}\,w_{i}^{2}}
{{\displaystyle\sum}\,W_{i}^{2}}\right)\,,
\end{equation}
where $N_{W}$ is the number of wavelet coefficients $W_{i}$ (and is equal to
the number of data points $N_{\mathrm{d}}$), and $n_{w}$ is the number of
wavelet coefficients $w_{i}$ that are not set to zero.  Clearly, there are
also visual criteria for judging the quality of the compressed data.

   The most important requirement for a good compression ability (large
$\mathit{CF}$ and small $\mathit{LI}$) is that the decomposition wavelet
should have many vanishing moments, and the basic reason is the following.  A
wavelet with $n$ vanishing moments is insensitive to polynomials of degree
$n-1$.  Regular data behave approximately as such polynomials in a
neighbourhood of a given point.  Hence the wavelet only feels the deviation
from such behaviour, which decreases with $n$.  Thus a large $n$ means that
the detail coefficients tend to be small and this implies a potentially good
compression ability.%
\footnote{In addition, it turns out that the first $n$ `multipole' moments of
          the data are conserved, starting from the zeroth-order one, if no
          approximation coefficient is set to zero.  This is particularly
          meaningful when the data represent a mass or a charge
          distribution.}
On the other hand, a good compression ability is guaranteed only if the
decomposition wavelet has a sufficiently small support.  Basically, the
analysis should be local enough otherwise the deviation from polynomial
behaviour of the data becomes significant.  (For a more explicit condition
see Sect.\ 4.2.)  Lastly, high regularity and symmetry are mainly needed by
the reconstruction wavelet for a good quality of the compressed data.  The
conclusion is that bi-orthogonal wavelets represent the best alternative for
satisfying the requirements above.

   Let us then consider a beautiful image of disc galaxies and choose an
appropriate wavelet pair, `rbio\,2.8', which belongs to the family of reverse
bi-orthogonal spline wavelets (see Daubechies 1992; here we use the same name
as in the code).  Fig.\ 2 illustrates the example \emph{eloquently}.

\begin{figure*}
\scalebox{.91}{\includegraphics{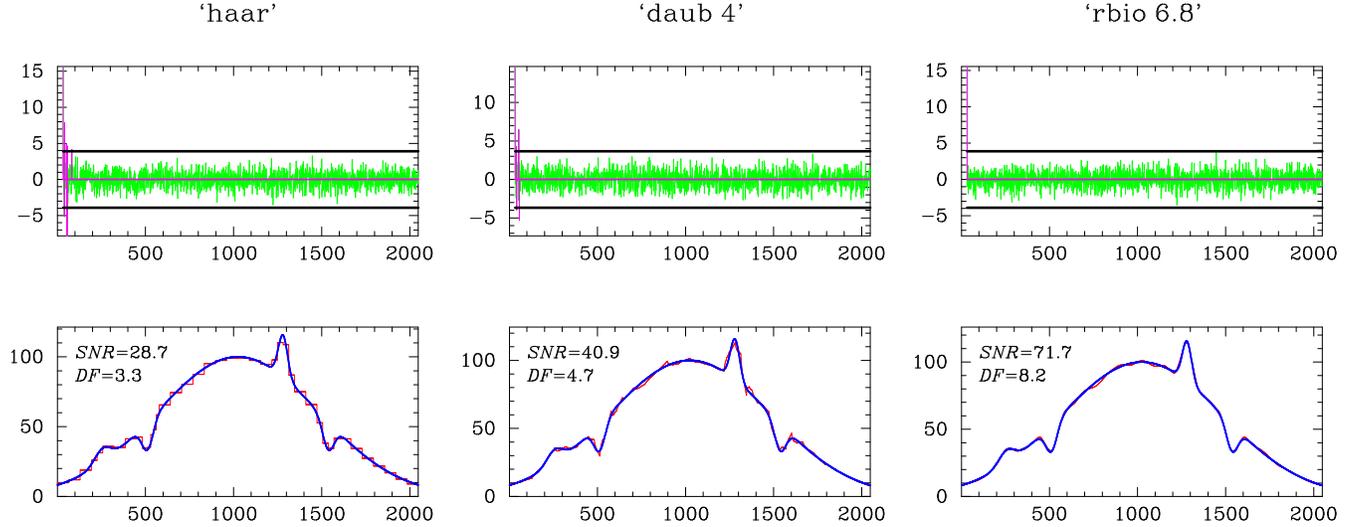}}
\caption{Original and thresholded wavelet coefficients (top), and de-noised
         data vs.\ the model data (bottom) for various choices of the
         wavelet.  Also specified are the signal-to-noise ratio
         $\mathit{SNR}$ in the de-noised data and the de-noising factor
         $\mathit{DF}$.  The reference case illustrated in Fig.\ 3
         corresponds to `bior\,4.4'.}
\end{figure*}

\subsection{De-noising: data and simulations}

The compression ability of the fast wavelet transform has a further important
implication: given noisy data, the underlying regular part gets mostly
concentrated into few large wavelet coefficients, whereas noise is mostly
mapped into many small wavelet coefficients.  In practice, this means that,
if we identify a correct threshold, then we can set all the small
coefficients to zero and get back data almost decontaminated from noise.
This is the idea behind data de-noising: a rigorous way to compress noisy
data.  In this section, we go on discussing the basics of de-noising.  A more
detailed discussion is given in Sect.\ 4.  The identification of a correct
threshold, which is crucial to the whole process of de-noising, is discussed
in Sect.\ 4.4.

   Let us now illustrate the basics of de-noising in a concrete case (cf.\
Fig.\ 3).  Fig.\ 3a shows data with Poissonian noise and the perfectly
non-noisy model data.%
\footnote{Poissonian data can be generated using the numerical recipes by
          Press et al.\ (1992).}
This type of noise is characterized by a multivariate Poissonian probability
distribution, and hence the local standard deviation of the data is equal to
the square root of their local mean:
$\sigma_{\mathrm{loc}}=\sqrt{\mu_{\mathrm{loc}}}$ (see, e.g., Bevington \&
Robinson 1992).  Poissonian noise occurs in all experiments and observations
where the data represent `counts' in a set of bins.  Fig.\ 3b shows that the
noisy data are pre-processed so as to transform Poissonian noise into
Gaussian white noise, where `white' means that it is equally significant on
all scales (constant power spectrum).  This type of noise is well known for
its mathematical tractability (see, e.g., Gonzalez \& Woods 2002).  In fact,
the reason for pre-processing the data is that Gaussian white noise is
convenient for identifying a correct threshold.  Figs 3c--f show the
remaining route after the choice of the wavelet: fast wavelet transforming
(Fig.\ 3c); thresholding the wavelet coefficients, which is the heart of
de-noising (Fig.\ 3d); inverse fast wavelet transforming (Fig.\ 3e); and,
finally, post-processing the data, which is needed after the initial
pre-processing (Fig.\ 3f).  The de-noised data are shown vs.\ the model data.

   As in Fig.\ 3 the model data are known, the de-noising ability can be
quantified by the signal-to-noise ratio $\mathit{SNR}$ and the de-noising
factor $\mathit{DF}$, defined as:
\begin{equation}
\mathit{SNR}=\left[\frac{{\displaystyle\sum}\,X_{i}^{2}}
{{\displaystyle\sum}\,(Y_{i}-X_{i})^{2}}\right]^{1/2}\,,
\end{equation}
\begin{equation}
\mathit{DF}=\frac{(\mathit{SNR})_{\mathrm{de}}}
{(\mathit{SNR})_{\mathrm{no}}}\,,
\end{equation}
where $X_{i}$ are the model data and $Y_{i}$ are either the noisy data (`no')
or the de-noised data (`de').  In addition, $(\mathit{SNR})_{\mathrm{de}}$
means the inverse of an appropriately defined estimation error.  Clearly,
there are also visual criteria for judging the quality of the de-noised data.
Fig.\ 3 illustrates the improvement produced by de-noising expressively.  In
general, the model data are not known so the de-noising ability and the
quality of the de-noised data are difficult to estimate.

   2-D or 3-D data de-noising is similar to the 1-D case.  The differences
are discussed together with other details of de-noising (see Sect.\ 4).

   \emph{How does de-noising work for N-body simulations?}  The de-noising
method discussed here applies to discrete data so it is natural to consider
grid-based $N$-body simulations.  Such simulations use a grid for tabulating
the particle density, and for computing the potential and the field (see,
e.g., Hockney \& Eastwood 1988).  The number of particles $n$ in each cell
shows fluctuations $|\delta n|/\langle n\rangle\sim\langle n\rangle^{-1/2}$
with respect to an average $\langle n\rangle$.  This means that the particle
distribution is polluted by noise that is basically Poissonian, whereas the
noise induced in the potential and in the field is of a more complex nature.
Using such a method we can thus de-noise the particle distribution at each
timestep and make the simulation equivalent to a simulation with many more
particles.  This is the idea behind our application (cf.\ Paper I).

\section{TOUR OF THE CODE}

As we have explained in Sect.\ 3.2, de-noising standard data and $N$-body
simulations consists of the following processes: pre-processing of the data,
choice of the wavelet, fast wavelet transform, thresholding of the wavelet
coefficients, inverse fast wavelet transform and post-processing of the data.
These are also the steps of the de-noising algorithm in our code.  In this
section, we discuss them in detail.  Advanced aspects of de-noising are
discussed in Sect.\ 5.

\subsection{Pre-processing of the data}

The data should be pre-processed if they are contaminated by Poissonian
noise.  The Poissonian data $Y_{\mathrm{P}}$ are transformed into data
$Y_{\mathrm{G}}$ with (additive) Gaussian white noise of standard deviation
$\sigma_{\mathrm{G}}=1$:
\begin{equation}
Y_{\mathrm{G}}=2\,\sqrt{Y_{\mathrm{P}}+{\textstyle\frac{3}{8}}}
\end{equation}
(Anscombe 1948), which can then be de-noised as discussed in Sects 4.2--4.5.
The Anscombe transformation has the remarkable property to help achieving
normalization, variance stabilization and additivity (see Stuart \& Ord
1991).  On the other hand, it has a tendency to fail locally where the data
have small values or large variations (e.g., Kolaczyk 1997; see also Starck,
Murtagh \& Bijaoui 1998).  On the whole, such an ingenious method produces
very good results if the data are post-processed appropriately (see Sect.\
4.6).

\begin{figure*}
\scalebox{.91}{\includegraphics{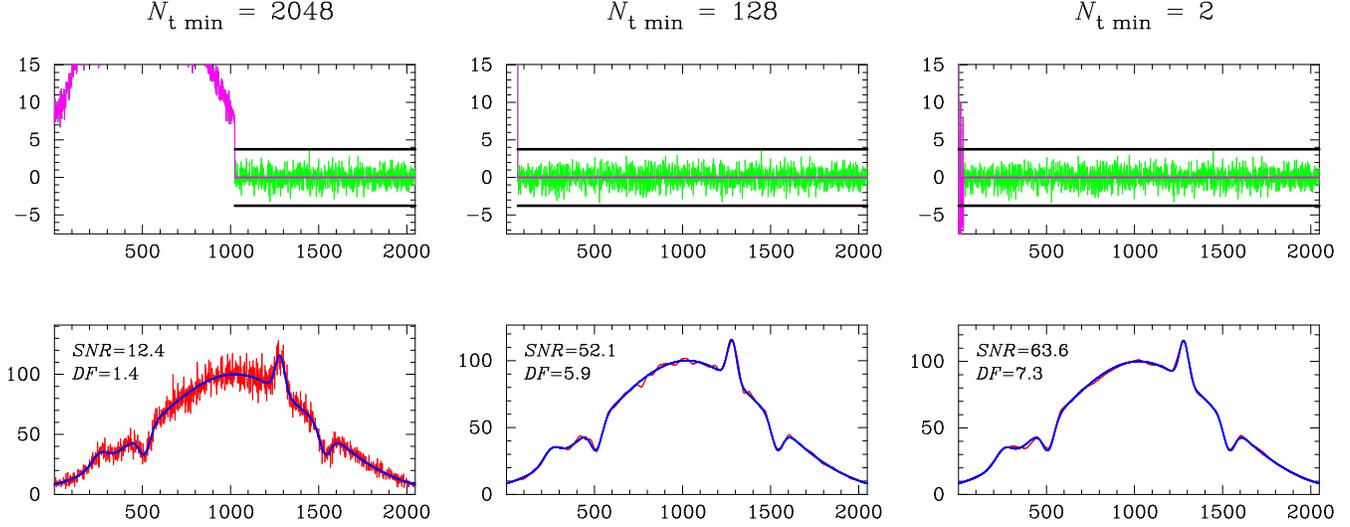}}
\caption{Original and thresholded wavelet coefficients (top), and de-noised
         data vs.\ the model data (bottom) for various values of the
         parameter $N_{\mathrm{t\,min}}$ defined in the text.  Also specified
         are the signal-to-noise ratio $\mathit{SNR}$ in the de-noised data
         and the de-noising factor $\mathit{DF}$.  The reference case
         illustrated in Fig.\ 3 corresponds to $N_{\mathrm{t\,min}}=64$.}
\end{figure*}

\subsection{Choice of the wavelet}

The wavelets included in the code belong to three families: the Daubechies
wavelets, the bi-orthogonal spline wavelets and the reverse bi-orthogonal
spline wavelets.  The last two families are intimately related: the
decomposition wavelets of the `reverse' family are the reconstruction
wavelets of the other, and vice versa.  Such wavelet families were introduced
by Daubechies (1992).  Various representatives of the wavelets included in
the code have already been shown in Fig.\ 1 and discussed in Sect.\ 2.4.  The
others have intermediate properties, or are the reverse of those illustrated.
The most useful wavelets of the code are specified at the end of this
section.

   How does the choice of the wavelet affect de-noising?  Let us consider the
representative wavelets mentioned above, since the others have intermediate
or similar effects.  The reference case illustrated in Fig.\ 3 corresponds to
`bior\,4.4', which we have chosen to discuss the basics of de-noising (see
Sect.\ 3.2).  Fig.\ 4 shows the effects of the other choices.  The wavelets
`haar' and `daub\,4' give rise to large irregular coefficients in the two
coarsest details, which exceed the threshold, and to small but significant
irregularities in the de-noised data.  The resulting signal-to-noise ratio
and de-noising factor are worse than in the reference case.  In contrast,
`rbio\,6.8' de-noises almost as well as `bior\,4.4'.

   Let us then explain the key points for a successful choice of the wavelet.
The conditions that should be fulfilled for a good de-noising are three.  (i)
The wavelet should satisfy the requirements for a good compression, which are
discussed in Sect.\ 3.1.  (ii) The wavelet should be orthogonal.
Orthogonality implies that Gaussian white noise in the data is transformed
into Gaussian white noise in the wavelet coefficients.  This is convenient
for a correct threshold identification, which is discussed in Sect.\ 4.4.
(iii) The size of the interval where the wavelet differs significantly from
zero should be comparable to the resolution needed by the data, or to the
effective spatial resolution of the simulations.  Such interval must not be
confused with the support of the wavelet.  Conditions (i)--(iii) cannot be
fulfilled equally well.  The best alternative is represented by bi-orthogonal
wavelet pairs that are also quasi-orthogonal, which are consistent with a
resolution of about three to four bin/mesh sizes.  The selected wavelets are:
`bior\,4.4' and `bior\,6.8', together with their reverse `rbio\,4.4' and
`rbio\,6.8' (cf.\ Fig.\ 1 and Sect.\ 2.4, and recall what `reverse' means).
In `bior\,$n_{1}.n_{2}$' the decomposition and reconstruction wavelets have
$n_{1}$ and $n_{2}$ vanishing moments, respectively; in `rbio\,$n_{1}.n_{2}$'
vice versa.  We cannot provide further reliable guide-lines on the most
appropriate choice of the wavelet.  It depends on the problem and can be
found through the optimization trial discussed in Sect.\ 6.

\begin{figure*}
\scalebox{1.}{\includegraphics{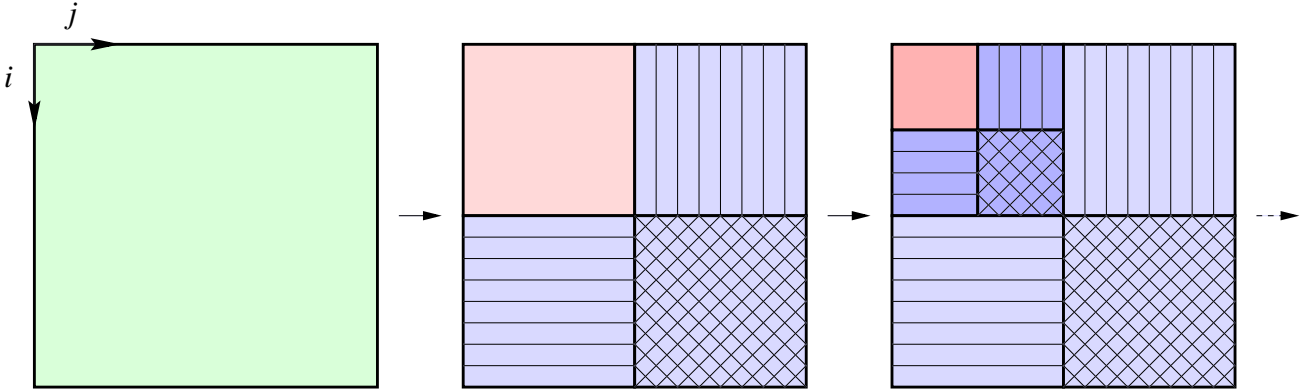}}
\caption{Action of the 2-D fast wavelet transform and structure of the
         transformed data.  The original data and the approximations are
         represented without patterns.  The details corresponding to
         vertical/horizontal/diagonal variations are represented by patterns
         of horizontal/vertical/diagonal lines (features along a certain
         direction have maximum gradient along the perpendicular direction).}
\end{figure*}

\subsection{Fast wavelet transform}

From the computational point of view, the choice of the wavelet corresponds
to the choice of a set of filters for the fast wavelet transform and its
inverse.  For a bi-orthogonal wavelet, they are: the high-pass and low-pass
decomposition filters $\tilde{g}_{i}$ and $\tilde{h}_{i}$, and the high-pass
and low-pass reconstruction filters $g_{i}$ and $h_{i}$, respectively.  In
the orthogonal case, $\tilde{g}_{i}=g_{i}$ and $\tilde{h}_{i}=h_{i}$.  The
coefficients of $\tilde{h}_{i}$ and $h_{i}$ are tabulated and centred as
close as possible to $i=0$, while those of $\tilde{g}_{i}$ and $g_{i}$ are
computed from the relations $\tilde{g}_{i+1}=(-1)^{i+1}h_{-i}$ and
$g_{i+1}=(-1)^{i+1}\tilde{h}_{-i}$.  The filters are padded with zeros so as
to be defined for $i=-M,\ldots,M$ ($M$ even), and to be consistent with the
formulae for the transforms.  In particular, `bior\,4.4' and `rbio\,4.4' have
$M=6$, while `bior\,6.8' and `rbio\,6.8' have $M=10$.  (The detailed
relations of such filters to the wavelets and scaling functions are very
complicated and irrelevant to our context; see Goedecker 1998.)

   One step of the (forward) fast wavelet transform replaces the current
approximation $A_{i}(N_{\mathrm{t}})$, of size $N_{\mathrm{t}}$, with a
coarser approximation $A_{i}(N_{\mathrm{t}}/2)$ and detail
$D_{i}(N_{\mathrm{t}}/2)$, of size $N_{\mathrm{t}}/2$:
\begin{equation}
A_{i}(N_{\mathrm{t}}/2)=\sum_{j=-M+1}^{M}\tilde{h}_{j}
A_{j+2i}(N_{\mathrm{t}})\,,
\end{equation}
\begin{equation}
D_{i}(N_{\mathrm{t}}/2)=\sum_{j=-M+1}^{M}\tilde{g}_{j}
A_{j+2i}(N_{\mathrm{t}})\,,
\end{equation}
where the index $j+2i$ is wrapped around when it gets out of the range
(periodic boundary conditions; see Goedecker 1998).  Initially,
$N_{\mathrm{t}}=N_{\mathrm{d}}$ and $A_{i}=X_{i}$, $X_{i}(N_{\mathrm{d}})$
being the original data.  The transform ends when
$N_{\mathrm{t}}=N_{\mathrm{t\,min}}$, so that the transformed data
$Y_{i}(N_{\mathrm{d}})$ consist of the coarse approximation
$A_{i}(N_{\mathrm{t\,min}}/2)$ and the sequence of finer and finer details
$D_{i}(N_{\mathrm{t\,min}}/2)$, $D_{i}(N_{\mathrm{t\,min}})$, \ldots,
$D_{i}(N_{\mathrm{d}}/2)$.  Note that $N_{\mathrm{t\,min}}$ is a free
parameter of the code.  If we assume that $N_{\mathrm{d}}$ is a power of 2,
then $N_{\mathrm{t\,min}}$ is also a power of 2 and such that $2\leq
N_{\mathrm{t\,min}}\leq N_{\mathrm{d}}$.  A complete transform corresponds to
$N_{\mathrm{t\,min}}=2$, but this value does not necessarily mean a good
de-noising.  A more general assumption is that $N_{\mathrm{d}}$ contains a
power of two, which has obvious implications for $N_{\mathrm{t\,min}}$.  Data
of different size can be padded (see Sect.\ 6).

   How does the value of the parameter $N_{\mathrm{t\,min}}$ affect
de-noising?  The reference case illustrated in Fig.\ 3 corresponds to
$N_{\mathrm{t\,min}}=64$ ($N_{\mathrm{d}}=2048$).  Fig.\ 5 shows the effects
of other values.  For $N_{\mathrm{t\,min}}=2048$, only the smallest-scale
noise is removed so the processed data are nearly as noisy as the original
data.  For $N_{\mathrm{t\,min}}=128$, there is residual noise on large
scales, and the resulting signal-to-noise ratio and de-noising factor are
worse than in the reference case.  For $N_{\mathrm{t\,min}}=2$, the
de-noising is complete and nearly as good as for $N_{\mathrm{t\,min}}=64$.
Nevertheless, there are large anomalous coefficients in the five coarsest
details, which exceed the threshold, and small anomalies in the de-noised
data.

   Let us then explain which values of $N_{\mathrm{t\,min}}$ imply a good
de-noising.  It must be $N_{\mathrm{t\,min}}\ll N_{\mathrm{d}}$, otherwise
de-noising is incomplete; and besides $N_{\mathrm{t\,min}}\ga4M$, otherwise
the wavelet becomes too dilated in comparison with the size of the data and
wrap-around effects become significant.  (Cosmological simulations are
peculiar in this context; see Sect.\ 5.3.)  The best value of
$N_{\mathrm{t\,min}}$ depends on the problem and can be found through the
optimization trial discussed in Sect.\ 6.

   Finally, we point out the (non-obvious) differences between the 2-D or 3-D
fast wavelet transform and the 1-D case.  Fig.\ 6 shows how the transform
acts on 2-D data.  In general, given $n$-D data of size $N_{\mathrm{d}}^{n}$,
the first step of the transform decomposes them into $2^{n}$ parts of size
$(N_{\mathrm{d}}/2)^{n}$: 1 approximation and $2^{n}-1$ details, one for each
axis and each diagonal.  It is done by 1-D transforming the data along each
index, for all values of the other indices, consecutively.  Then the
discussion basically follows the 1-D case, except that the complexity of the
transform increases by a factor of $n\,2^{n-1}/(2^{n}-1)$ with respect to
$4MN_{\mathrm{d}}^{n}$.  The generalization to data of size
$N_{\mathrm{d}1}\cdots N_{\mathrm{d}n}$ is plain.

\begin{figure*}
\scalebox{.91}{\includegraphics{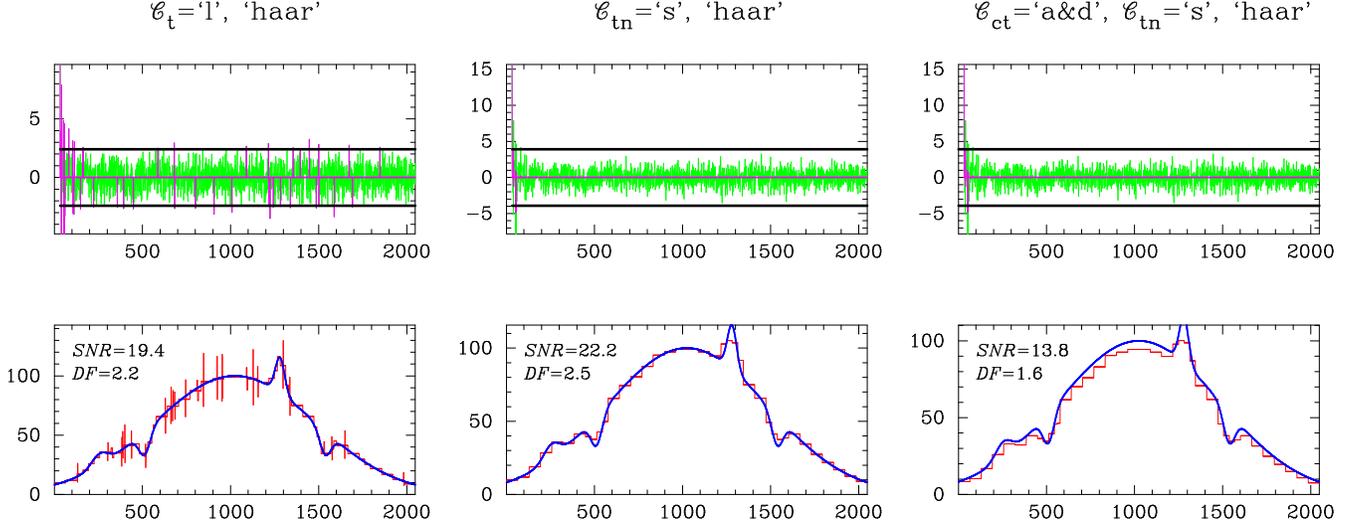}}
\caption{Original and thresholded wavelet coefficients (top), and de-noised
         data vs.\ the model data (bottom) for various thresholding options
         and the `haar' wavelet.  Also specified are the signal-to-noise
         ratio $\mathit{SNR}$ in the de-noised data and the de-noising factor
         $\mathit{DF}$.  The reference case illustrated in Fig.\ 3
         corresponds to $\mathcal{C}_{\mathrm{t}}=\mbox{`h'}$,
         $\mathcal{C}_{\mathrm{ct}}=\mbox{`d'}$,
         $\mathcal{C}_{\mathrm{tn}}=\mbox{`h'}$ and the `bior\,4.4' wavelet
         [for the corresponding `haar' case cf.\ Fig.\ 4 (left)].}
\end{figure*}

\subsection{Thresholding of the wavelet coefficients}

The heart of de-noising consists of identifying a correct threshold, and
deciding which type of wavelet coefficients are to be thresholded and how.
Note the difference between thresholding and smoothing, where the detail
coefficients below a given scale are set to zero independent of their value.
In the following, we discuss thresholding and introduce the options of the
code.

   As pointed out in Sect.\ 4.2, a correct threshold can be identified if the
wavelet is orthogonal, or quasi-orthogonal.  The threshold $T$ is
proportional to the standard deviation of noise $\sigma$, and the
proportionality factor $K$ depends on the size of the data:
\begin{equation}
T=K(N_{\mathrm{d}})\,\sigma\,.
\end{equation}
If the \underline{s}tandard \underline{d}eviation is \underline{n}ot
\underline{g}iven ($\mathcal{C}_{\mathrm{sd}}=\mbox{`ng'}$), as in the case
of Gaussian white noise, then it is robustly estimated through the median
absolute deviation of the finest detail:
\begin{equation}
\sigma\simeq\frac{1}{0.6745}\,\mathit{MAD}[D_{i}(N_{\mathrm{d}}/2)]\,.
\end{equation}
A robust estimator and the finest detail are used for minimizing the
contribution of outlying wavelet coefficients, which are not caused by noise.
If the \underline{s}tandard \underline{d}eviation is \underline{g}iven
($\mathcal{C}_{\mathrm{sd}}=\mbox{`g'}$), as in the case of Poissonian noise,
then
\begin{equation}
\sigma\simeq1\,.
\end{equation}
Concerning $K(N_{\mathrm{d}})$, it is rigorously determined so that the
threshold matches both the noise level and the significance level of the
wavelet coefficients, according to probability criteria.  We can decide
between two functional forms for $K(N_{\mathrm{d}})$.  One corresponds to a
\underline{h}igher \underline{t}hreshold
($\mathcal{C}_{\mathrm{t}}=\mbox{`h'}$), which is more effective but less
safe:
\begin{equation}
K(N_{\mathrm{d}})=\sqrt{2\ln N_{\mathrm{d}}}\,.
\end{equation}
The other corresponds to a \underline{l}ower \underline{t}hreshold
($\mathcal{C}_{\mathrm{t}}=\mbox{`l'}$), and is approximated analytically as:
\begin{equation}
K(N_{\mathrm{d}})\simeq\left\{
 \begin{array}{ll}
 0                                   & \mbox{if\ }N_{\mathrm{d}}\leq32\,, \\
 0.3936+0.1829\log_{2}N_{\mathrm{d}} & \mbox{else}\,.
 \end{array}
\right.
\end{equation}
Next, the wavelet \underline{c}oefficients to \underline{t}hreshold can be
either the \underline{d}etails ($\mathcal{C}_{\mathrm{ct}}=\mbox{`d'}$):
\begin{equation}
W_{i}=D_{i}(N_{\mathrm{t\,min}}/2),\ldots,D_{i}(N_{\mathrm{d}}/2)\,;
\end{equation}
or the \underline{a}pproximation \underline{\&} the \underline{d}etails
($\mathcal{C}_{\mathrm{ct}}=\mbox{`a\&d'}$):
\begin{equation}
W_{i}=A_{i}(N_{\mathrm{t\,min}}/2),D_{i}(N_{\mathrm{t\,min}}/2),\ldots,
D_{i}(N_{\mathrm{d}}/2)\,.
\end{equation}
The last option concerns the \underline{t}hresholding (method),
\underline{n}amed as in the literature.  It can be either \underline{h}ard
($\mathcal{C}_{\mathrm{tn}}=\mbox{`h'}$):
\begin{equation}
\overline{W}_{i}=\left\{
 \begin{array}{ll}
 0     & \mbox{if\ }|W_{i}|\leq T\,, \\
 W_{i} & \mbox{else}\,;
 \end{array}
\right.
\end{equation}
or \underline{s}oft ($\mathcal{C}_{\mathrm{tn}}=\mbox{`s'}$):
\begin{equation}
\overline{W}_{i}=\left\{
 \begin{array}{ll}
 0                               & \mbox{if\ }|W_{i}|\leq T\,, \\
 \mathrm{sign}(W_{i})(|W_{i}|-T) & \mbox{else}\,.
 \end{array}
\right.
\end{equation}
So $\mathcal{C}_{\mathrm{tn}}=\mbox{`s'}$ means that even the wavelet
coefficients above $T$ are thresholded, and this is done shrinking them by
$T$.

   How do the thresholding options affect de-noising?  The reference case
illustrated in Fig.\ 3 corresponds to $\mathcal{C}_{\mathrm{t}}=\mbox{`h'}$,
$\mathcal{C}_{\mathrm{ct}}=\mbox{`d'}$ and
$\mathcal{C}_{\mathrm{tn}}=\mbox{`h'}$
($\mathcal{C}_{\mathrm{sd}}=\mbox{`ng'}$).  Fig.\ 7 shows the effects of
other options.  As the situation is degenerate, we consider the `haar'
wavelet instead of `bior\,4.4', since it reduces the degeneracy and its
effects have already been shown [cf.\ Fig.\ 4 (left)].  For
$\mathcal{C}_{\mathrm{t}}=\mbox{`l'}$, the threshold is exceeded by several
noisy detail coefficients, which give rise to spikes in the de-noised data.
For $\mathcal{C}_{\mathrm{tn}}=\mbox{`s'}$, the soft thresholding of the
detail coefficients over-regularizes the de-noised data, softening the maxima
and minima.  If in addition $\mathcal{C}_{\mathrm{ct}}=\mbox{`a\&d'}$, then
the thresholding of the approximation coefficients biases the de-noised data.

   Let us then explain which thresholding options imply a good de-noising.
We suggest the more effective option $\mathcal{C}_{\mathrm{t}}=\mbox{`h'}$
for standard data, unless they are expected to have meaningful irregularities
below the maximum noise level; while we recommend the safer option
$\mathcal{C}_{\mathrm{t}}=\mbox{`l'}$ for simulations.  In addition, it must
be $\mathcal{C}_{\mathrm{ct}}=\mbox{`d'}$ and
$\mathcal{C}_{\mathrm{tn}}=\mbox{`h'}$, otherwise the de-noised data get
biased and over-regularized, respectively.  (In advanced de-noising,
$\mathcal{C}_{\mathrm{ct}}$ and $\mathcal{C}_{\mathrm{tn}}$ are replaced by a
more useful parameter; see Sect.\ 5.1.)  Finally, it is natural to opt for
$\mathcal{C}_{\mathrm{sd}}=\mbox{`ng'}$ in the case of Gaussian white noise,
and for $\mathcal{C}_{\mathrm{sd}}=\mbox{`g'}$ in the case of Poissonian
noise.  If the Poissonian data have a sufficiently high signal-to-noise
ratio, such that the estimated $\sigma\simeq1$, then
$\mathcal{C}_{\mathrm{sd}}=\mbox{`ng'}$ fine-tunes the accuracy of
de-noising.  The algorithm gets slightly slower since the computation of the
median has complexity $O(N_{\mathrm{d}})$.  Therefore
$\mathcal{C}_{\mathrm{sd}}=\mbox{`ng'}$ may be a better option than
$\mathcal{C}_{\mathrm{sd}}=\mbox{`g'}$ for standard data, not for simulations
(where the signal-to-noise ratio is low and speed is a primary factor).

   Finally, thresholding in 2D or 3D is similar to the 1-D case and the
generalization to data of size $N_{\mathrm{d}1}\cdots N_{\mathrm{d}n}$ is
plain, except for the more complicated structure of the wavelet coefficients.

\begin{figure*}
\scalebox{.91}{\includegraphics{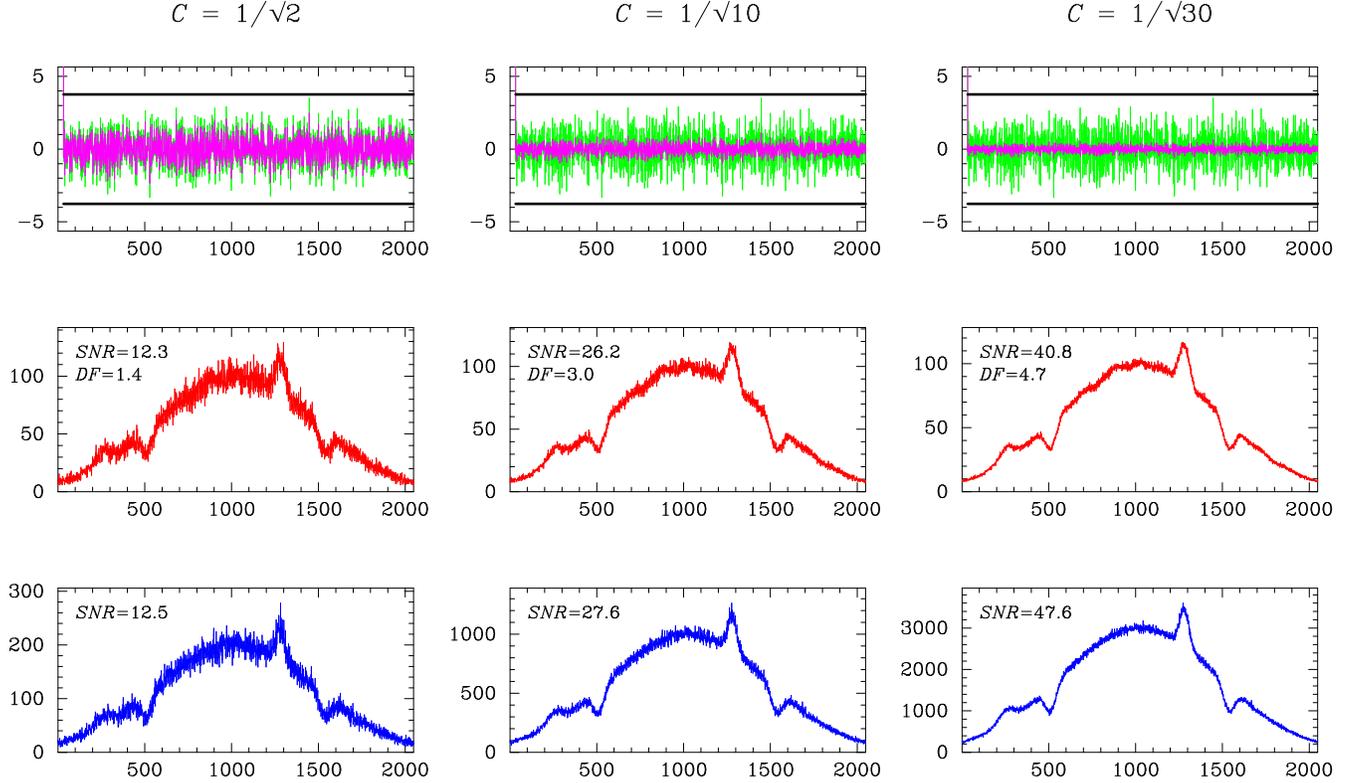}}
\caption{Partial de-noising at a pre-assigned level.  Original and
         thresholded wavelet coefficients (top), and de-noised data (middle)
         for various values of the contraction parameter $C$; also shown for
         comparison are realizations of Poissonian data with $1/C^{2}$ times
         more `counts' than in the original noisy data (bottom).  The
         signal-to-noise ratio $\mathit{SNR}$ and the de-noising factor
         $\mathit{DF}$ are specified.  The reference case illustrated in
         Fig.\ 3 corresponds to $C=0$ (total de-noising).  The implication
         for simulations is explained in the text.}
\end{figure*}

\subsection{Inverse fast wavelet transform}

One step of the inverse (backward) fast wavelet transform replaces the
current approximation $A_{i}(N_{\mathrm{t}}/2)$ and coarsest detail
$D_{i}(N_{\mathrm{t}}/2)$ with a finer approximation $A_{i}(N_{\mathrm{t}})$:
\begin{equation}
A_{2i}(N_{\mathrm{t}})=\!\!\!\!\!\sum_{j=-M/2}^{M/2-1}\!\!\!\!\!h_{2j}
A_{i-j}(N_{\mathrm{t}}/2)+g_{2j}D_{i-j}(N_{\mathrm{t}}/2)\,,
\end{equation}
\begin{equation}
A_{2i+1}(N_{\mathrm{t}})=\!\!\!\!\!\sum_{j=-M/2}^{M/2-1}\!\!\!\!\!h_{2j+1}
A_{i-j}(N_{\mathrm{t}}/2)+g_{2j+1}D_{i-j}(N_{\mathrm{t}}/2)\,,
\end{equation}
where the index $i-j$ is wrapped around when it gets out of the range (see
Goedecker 1998), $N_{\mathrm{t}}$ goes from $N_{\mathrm{t\,min}}$ to
$N_{\mathrm{d}}$, and so on (see Sect.\ 4.3).

   The inverse fast wavelet transform can be used for plotting wavelets, as
in Fig.\ 1.  Consider the data $X_{i}=\delta_{in}$ and the inverse
transformed data $Y_{i}$.  A discrete approximation of
$\psi_{\mathrm{rec}}(x)$ or $\phi_{\mathrm{rec}}(x)$ can be computed from
$Y_{i}$ through the following operations: scaling, translation, normalization
and possibly wrap-around.  The accuracy of the approximation, the type of
function and the parameters of the operations depend on $N_{\mathrm{d}}$,
$N_{\mathrm{t\,min}}$ and $n$.  Including the reverse set of filters produces
$\psi_{\mathrm{dec}}(x)$ or $\phi_{\mathrm{dec}}(x)$.  In Fig.\ 1, we have
set: $N_{\mathrm{d}}=2^{14}$, $N_{\mathrm{t\,min}}=2^{7}$,
$n=3N_{\mathrm{t\,min}}/4$ for the wavelet plots and
$n=N_{\mathrm{t\,min}}/4$ for the scaling-function plots.  In both plots, the
discrete argument is $x_{i}=(i-b)/a$ and the function is
$f(x_{i})=\sqrt{a}\,Y_{i}$, with $a=2N_{\mathrm{d}}/N_{\mathrm{t\,min}}$ and
$b=N_{\mathrm{d}}/2$.  Such a set-up provides an accurate approximation and
avoids wrap-around, and thus it can be used for plotting wavelets in general.

\subsection{Post-processing of the data}

The data should be post-processed in the case of Poissonian noise.
Post-processing consists of inverse Anscombe transforming plus two
corrections.  The first correction is needed if the Anscombe transformation
fails locally (see Sect.\ 4.1), giving rise to small negative values in the
de-noised data.  We can correct such values by setting them to zero.  The
second correction is required because the Anscombe transformation introduces
a local bias in the data.  That is, if $\mu_{\mathrm{P}}$ is the local mean
of $Y_{\mathrm{P}}$ and $\mu_{\mathrm{G}}$ is the local mean of
$Y_{\mathrm{G}}$, the mean $\mu_{\mathrm{P}}^{\prime}$ estimated by inverse
transforming $\mu_{\mathrm{G}}$ is not equal to $\mu_{\mathrm{P}}$, and their
difference is the local bias of the transformation.  Starck et al.\ (1998)
have implied that the bias is multiplicative and unbounded, while Kolaczyk
(1997) has implied that the bias is additive and bounded but has not
estimated it.  Indeed, the comprehensive book by Stuart \& Ord (1991) shows
that the bias of the Anscombe transformation is additive and bounded, and can
be estimated analytically:
\begin{equation}
\mathit{BIAS}\simeq-\frac{1}{4}\left(1-\frac{1}{N_{\mathrm{d}}}\right)
\sigma_{\mathrm{G}}^{2}\,.
\end{equation}
This means that, with very little effort, we can subtract the bias almost
completely from the de-noised data.  And, if even a slight global bias is
unacceptable, then we can compute it numerically and subtract it completely
from the de-noised data.

\section{EXPLORING ADVANCED DE-NOISING}

Consider a simulation where the particle distribution is denoised at each
timestep.  If initially the de-noising factor is $\mathit{DF}_{0}$, then the
de-noised model has the same signal-to-noise ratio as a noisy model with
$\mathit{DF}_{0}^{2}$ times more particles.  This follows from the fact that
the noise is basically Poissonian, and hence the signal-to-noise ratio is
proportional to the square root of the average number of particles per cell.
Besides, as $\mathit{DF}$ depends more on the de-noising ability of the
wavelet method than on the characteristics of the particle distribution, we
can draw a more general conclusion: the de-noised simulation itself is
roughly equivalent to a noisy simulation with $\mathit{DF}_{0}^{2}$ times
more particles (cf.\ Paper I).  Until now we have learned how to de-noise so
as to get the largest $\mathit{DF}$ (see Sects 3.2 and 4).  On the other
hand, in simulations we do not always want to suppress noise totally.  We may
instead want to reduce it partially in order to understand and control its
effects.  In this section, we learn how to carry out such an advanced
de-noising.

\subsection{Partial de-noising at a pre-assigned level}

\begin{figure}
\centering
\scalebox{1.}{\includegraphics{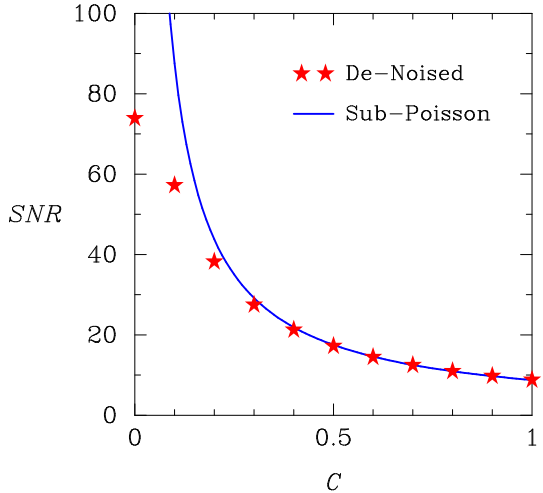}}
\caption{Accuracy of partial de-noising at a pre-assigned level.  The
         signal-to-noise ratio $\mathit{SNR}$ is shown as a function of the
         contraction parameter $C$ for the de-noised data and for
         sub-Poissonian data.  The sub-Poissonian data have local standard
         deviation $\sigma_{\mathrm{sub}}=C\sqrt{\mu_{\mathrm{sub}}}$ and
         local mean $\mu_{\mathrm{sub}}=\mu_{\mathrm{P}}$, where
         $\mu_{\mathrm{P}}$ refers to the original data with Poissonian
         noise; hence
         $(\mathit{SNR})_{\mathrm{sub}}=(\mathit{SNR})_{\mathrm{P}}/C$.  The
         accuracy is better than 10\% for $C\ga1/\sqrt{20}$, and it gets
         worse than 20\% for $C\la1/\sqrt{40}$.  The implication for
         simulations is explained in the text.}
\end{figure}

\subsubsection{Method and implementation}

Can we de-noise a simulation so as to make it equivalent to a simulation with
$\Gamma$ times more particles, for a pre-assigned level $\Gamma$?  Yes!  And
the idea is the following.  Recall what hard thresholding of the details
means (see Sect.\ 4.4), and consider the wavelet coefficients below the
threshold.  If we contract them by $C$, instead of setting them to zero, then
the noise level decreases by the same factor whereas the `signal' does not
change.  Hence the signal-to-noise ratio increases by a factor of $1/C$, and
the simulation becomes equivalent to a simulation with $1/C^{2}$ times more
particles.  Thus the problem is solved if we set $C=1/\sqrt{\Gamma}$.  (For
an analogous thresholding in the context of speech signals see Storm 1998.)

   We now illustrate this idea in the simple, but instructive, context of
standard data.  Fig.\ 8 shows that partial de-noising at a pre-assigned level
works as expected.  Note that this type of de-noising is meant to turn data
with Poissonian noise into `sub-Poissonian' data.  In such data the original
Poissonian deviations from the local mean are contracted by $C$, while
obviously the de-noised data are subject to an estimation error.  Fig.\ 9
shows that, as expected, the accuracy of partial de-noising at a pre-assigned
level is very good except for $C\la2/(\mathit{DF})_{\mathrm{tot}}$, where
$(\mathit{DF})_{\mathrm{tot}}$ refers to the case of total de-noising
($C=0$).

\begin{figure*}
\scalebox{.96}{\includegraphics{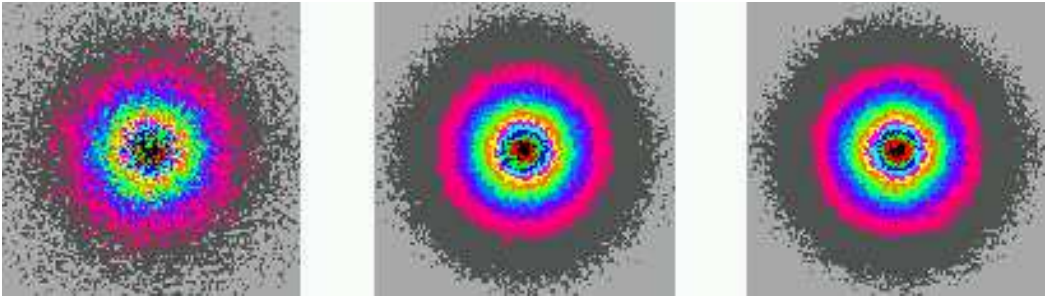}}
\caption{Partial de-noising at a pre-assigned level in action: comparison
         between the initial models.  The set of simulation models has a
         physical grid of $N_{\mathrm{c}}=N_{\mathrm{d}}^{2}=256^{2}$ cells
         and cell size $\Delta_{\mathrm{c}}=0.25$ kpc.  The particle
         distribution is shown for the noisy model with $N=10^{5}$ particles
         (left), the noisy model with $N=10^{6}$ (middle), and the de-noised
         model with $N=10^{5}$ and contraction parameter $C=1/\sqrt{10}$
         (right).  In each model the signal-to-noise ratio is
         $\mathit{SNR}\simeq5.7,17.8,17.1$, respectively.  As expected, the
         accuracy of partial de-noising at a pre-assigned level is very good
         for such initial models.}
\end{figure*}

\begin{figure*}
\scalebox{.96}{\includegraphics{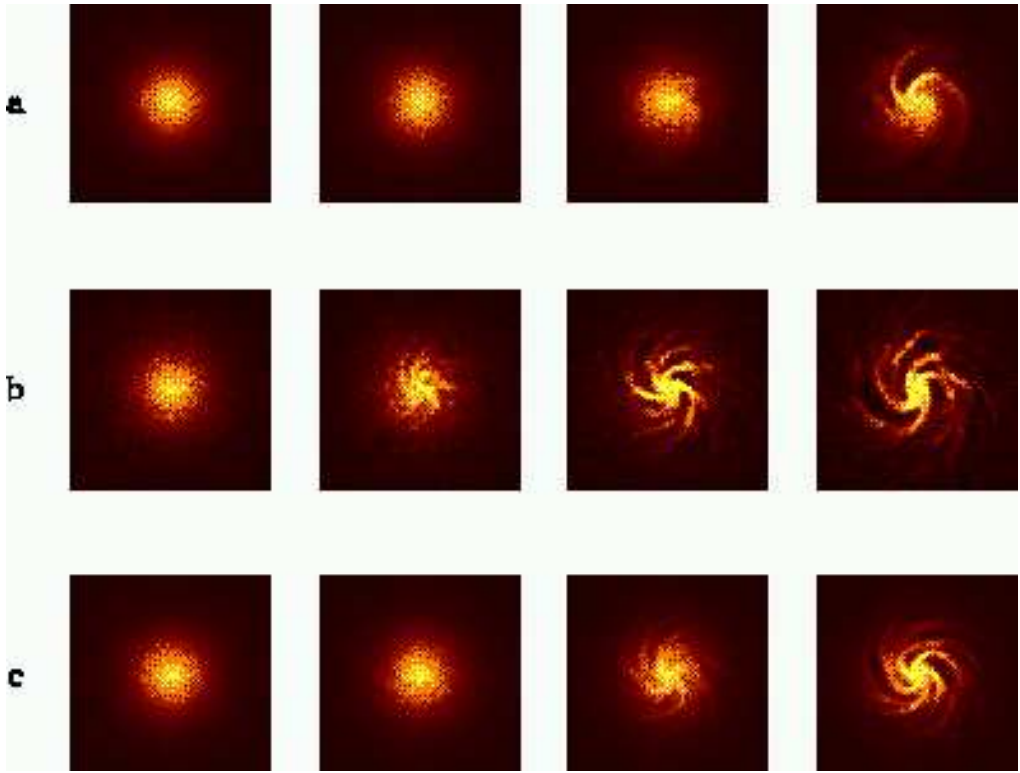}}
\caption{Partial de-noising at a pre-assigned level in action: fragmentation
         of a cool galactic disc.  \textbf{a}: de-noised simulation with
         $N=10^{5}$ and $C=1/\sqrt{10}$; \textbf{b}: noisy simulation with
         $N=10^{5}$; \textbf{c}: noisy simulation with $N=10^{6}$.  The
         initial models are the same as in Fig.\ 10.  For each simulation,
         the particle distribution is shown from 0 Myr to 150 Myr at
         intervals of 50 Myr (from left to right).  The time $\tau$ at which
         the initial axial symmetry breaks is a measure of the effect of
         noise on the simulation: a long $\tau$ means a weak effect.  As
         expected, $\tau$ increases from \textbf{b} to \textbf{c}; we also
         notice that $\tau$ is a little longer in \textbf{a} than in
         \textbf{c}.}
\end{figure*}

\subsubsection{Bench-marks}

Let us then explore this idea in simulations of disc galaxies.  The
examination is based on four bench-marks, originally introduced in Paper I.
\begin{enumerate}
   \item The first natural bench-mark is the comparison between the initial
models.
   \item The second bench-mark concerns the fragmentation of a cool galactic
disc, which is the onset of a gravitational instability (see, e.g., Binney \&
Tremaine 1987).  A rotating disc with low velocity dispersion is
gravitationally unstable and therefore sensitive to perturbations, which are
amplified and break the initial axial symmetry of the system (e.g., Semelin
\& Combes 2000; Huber \& Pfenniger 2001).  The time that characterizes
symmetry breaking clearly depends on the initial amplitude of the
perturbations, for small perturbations need a long time to grow into an
observable level.  In particular, this is true for the fluctuations imposed
by granular initial conditions.  Thus the symmetry-breaking time is a clear
diagnostic for quantifying the effect of noise on the simulation.
   \item The third bench-mark concerns the heating following the
fragmentation.  This is a fundamental process in the dynamical evolution of
disc galaxies, which is induced by gravitational instabilities via the
outward transport of angular momentum and energy (see, e.g., Binney \&
Tremaine 1987).  Therefore this bench-mark has a clear physical motivation.
When spiral gravitational instabilities reach a sufficiently large amplitude,
the velocity dispersion of the disc starts to increase by collective
relaxation (e.g., Zhang 1998; Griv, Gedalin \& Yuan 2002).  The heat produced
in a dynamical time is low if the initial amplitude of the instabilities is
small.  Thus the increase of velocity dispersion is another diagnostic for
quantifying the effect of noise on the simulation.
   \item The fourth bench-mark concerns the accretion following the
fragmentation.  This is also a fundamental process in the dynamical evolution
of disc galaxies (see, e.g., Binney \& Tremaine 1987).  Therefore this
bench-mark also has a clear physical motivation.  The amplification of spiral
gravitational instabilities produces not only heating but also
re-distribution of matter in the disc, which appears more evidently as
accretion near the centre (e.g., Zhang 1998; Griv et al.\ 2002).  The mass
accreted in a dynamical time is low if the initial amplitude of the
instabilities is small.  Thus the peak of mass density is still another
diagnostic for quantifying the effect of noise on the simulation.
\end{enumerate}
We consider the same basic simulation as in Paper I, which has $N=10^{5}$
particles.  We de-noise it choosing the `rbio\,6.8' wavelet, and setting
$N_{\mathrm{t\,min}}=16$ and $C=1/\sqrt{10}$, so as to make it equivalent to
a simulation with 10 times more particles (the thresholding options are the
usual ones for simulations; see Sect.\ 4.4).  The conservation of angular
momentum and energy is not significantly affected.  In fact, the deviations
are less than 0.02\% and 0.04\% per dynamical time, respectively, and compare
well with those typical of the code (Combes et al.\ 1990).  We also run the
noisy simulation with $N=10^{6}$.  This suite of simulations is sufficient
for the present purpose.  For further comparison see the extensive survey
presented in Paper I (the de-noised simulation has $C=0$).

   Figs 10--13 illustrate that partial de-noising at a pre-assigned level
works as expected, and the agreement is very good, except that the de-noised
simulation takes a little longer time to form the initial transient
structures (cf.\ Fig.\ 11).  The reason for this imperfection is twofold: it
concerns the de-noising itself (threshold) and the initial conditions (noisy
starts).
\begin{description}
   \item \emph{Threshold.}  At the beginning of the simulation, there is no
way to differentiate instabilities from amplified noisy fluctuations.
Thresholding weakens the initial instabilities until the relevant wavelet
coefficients exceed the threshold.  The usual threshold tends to be slightly
too high, and therefore the onset of the initial instabilities is delayed.
   \item \emph{Noisy starts.}  We know that the initial particle positions
and velocities are noisy.  We also know that the particle density is
partially de-noised, and the computed field has a consistent noise level.
This means that the excess of noise remains confined in phase space and does
not propagate dynamically.  In fact, in the de-noised simulation the
amplitude of the statistical fluctuations is similar to the noisier case,
whereas the evolution of the relevant quantities is similar to the less noisy
case (cf.\ Figs 12 and 13).  On the other hand, the excess of noise makes
instabilities less coherent, and therefore it delays their onset.
\end{description}

   Thus it is hard to evaluate the accuracy of partial de-noising at a
pre-assigned level for simulations, even if the accuracy is better than a few
per cent for the initial models (cf.\ Fig.\ 10).  But the conclusion is
strong anyway: our method, and code, can be used for understanding and
controlling the effects of noise on simulations.

\begin{figure*}
\scalebox{.96}{\includegraphics{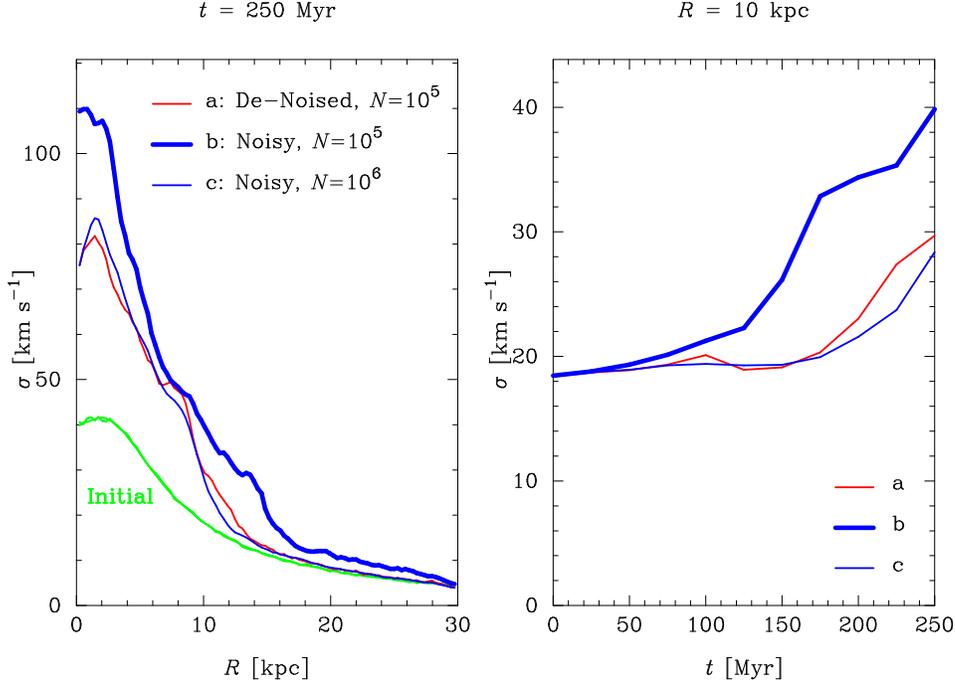}}
\caption{Partial de-noising at a pre-assigned level in action: heating
         following the fragmentation of a cool galactic disc.  The
         simulations are the same as in Fig.\ 11.  The velocity dispersion
         $\sigma$ is shown as a function of radius $R$ at the initial and
         final times, and as a function of time $t$ at an intermediate
         radius.  The increase of velocity dispersion $\Delta\sigma(R)$ from
         the initial to the final value is a measure of the effect of noise
         on the simulation: a small $\Delta\sigma$ means a weak effect.  In
         the simulations, except b, heating is significant only for $R\la12$
         kpc.  As expected,
         $\Delta\sigma_{\mathrm{b}}>\Delta\sigma_{\mathrm{c}}$ and
         $\Delta\sigma_{\mathrm{a}}\approx\Delta\sigma_{\mathrm{c}}$.}
\end{figure*}

\begin{figure}
\centering
\scalebox{.96}{\includegraphics{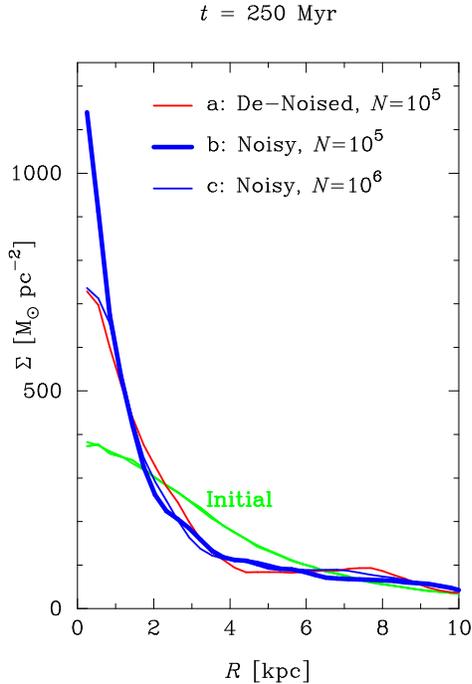}}
\caption{Partial de-noising at a pre-assigned level in action: accretion
         following the fragmentation of a cool galactic disc.  The
         simulations are the same as in Fig.\ 11.  The mass density $\Sigma$
         is shown as a function of radius $R$ at the initial and final times.
         The peak of final mass density $\hat{\Sigma}$ near the centre is a
         measure of the effect of noise on the simulation: a low
         $\hat{\Sigma}$ means a weak effect.  As expected,
         $\hat{\Sigma}_{\mathrm{b}}>\hat{\Sigma}_{\mathrm{c}}$ and
         $\hat{\Sigma}_{\mathrm{a}}\approx\hat{\Sigma}_{\mathrm{c}}$.}
\end{figure}

\subsubsection{Controlling the effectiveness of de-noising}

Before exploring more advanced aspects of de-noising, let us reflect on the
main differences between partial de-noising at a pre-assigned level and total
de-noising, and explain what we mean by noise control (or analogous terms).
Consider a simulation with, say, $N=10^{5}$ particles.  And suppose that we
decide beforehand to make it equivalent to a simulation with, say,
$N_{\mathrm{pre}}=1.8\times10^{6}$ particles (pre-assigned number).  Then we
set the contraction parameter $C=1/\sqrt{18}$, and run the simulation.  The
partially de-noised model will accurately mimic a noisy model with
$\Gamma=18$ times more particles.  This is noise control.  On the other hand,
the accuracy of partial de-noising at a pre-assigned level deteriorates for
small $C$ (cf.\ Fig.\ 9 and its discussion).  In the limit $C\rightarrow0$
the effectiveness of de-noising becomes maximum, but we cannot predict it
with sufficient accuracy unless we compare the initial models quantitatively
(see Paper I).  If the method were perfect we would have an improvement in
the equivalent number of particles by a factor of $1/C^{2}\rightarrow\infty$,
whereas in practice the improvement is by a factor of about 100 (cf.\ Paper
I).  In such a case we do not have control over noise, but we exploit total
(maximum) de-noising.

\begin{figure*}
\scalebox{.85}{\includegraphics{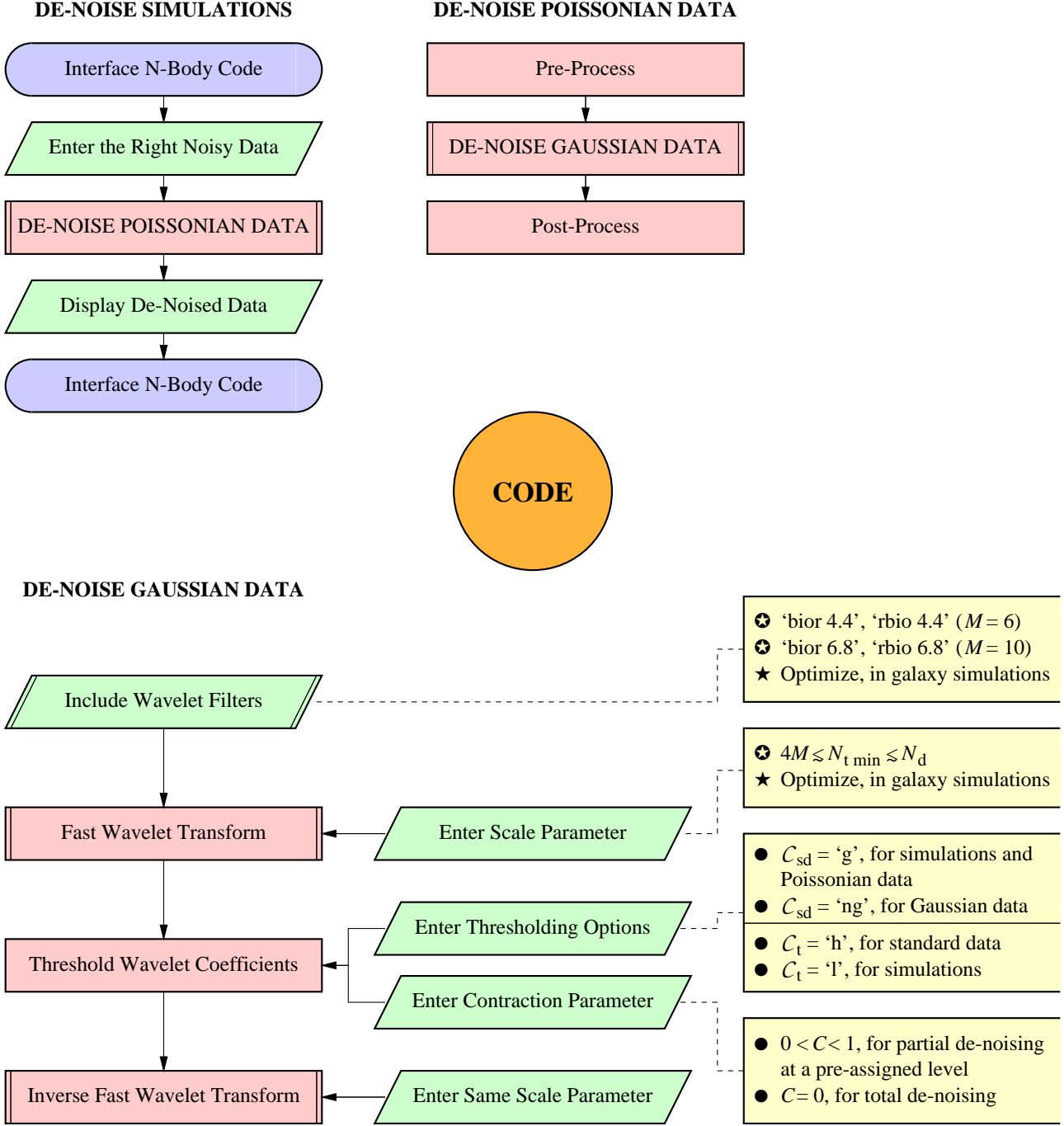}}
\caption{Flowchart of the code.}
\end{figure*}

\subsection{Partially noisy starts and adaptive de-noising}

Can we achieve an even better noise control?  Yes, in principle, and the idea
is the following.  We should first impose appropriate initial conditions so
as to make the model equivalent to a model with $\Gamma$ times more
particles, without de-noising it.  Such partially noisy starts can be
generated by setting up a fraction $1/\Gamma$ of the particles with noisy
starts, and the rest with quiet starts (quiet starts are common in plasma
simulations; see, e.g., Dawson 1983; Birdsall \& Langdon 1991).  Doing so,
the initial particle distribution is basically sub-Poissonian and its
signal-to-noise ratio is higher than in the Poissonian case by a factor of
$\sqrt{\Gamma}$.  On the other hand, the noise has a natural tendency to
become fully Poissonian in few dynamical times, as a reaction of the system
to the imposed order and hence reduced entropy.  We should then de-noise the
simulation consistently.  At the initial time, the threshold $T$ is lower
than the usual one by a factor of $1/\sqrt{\Gamma}$ and the contraction
parameter $C$ is unity (no de-noising).  When the noise level increases, $T$
should be increased accordingly and $C$ should be decreased in inverse
proportion (in the case of quiet starts, $C=0$; see also Paper I).  Such an
adaptive de-noising is not yet implemented in the code.

\subsection{Partial de-noising up to a given scale}

Cosmological simulations are peculiar with respect to galaxy and plasma
simulations.  The initial conditions consist of setting up a quiet uniform
particle distribution, and of imposing small random fluctuations with
Gaussian statistics and a given power spectrum (e.g., Efstathiou et al.\
1985; Sylos Labini et al.\ 2003).  Such fluctuations are amplified by
gravitational instability and form stuctures.  On the other hand, Poissonian
noise develops on the same time-scale and therefore affects structure
formation.  The onset of Poissonian noise is especially quick in cold dark
matter simulations, where structures form bottom-up and the first virialized
systems contain a small number of particles (e.g., Binney \& Knebe 2002;
Diemand et al.\ 2004).

   Thus de-noising cosmological simulations is a very complex and demanding
problem: the method should remove noise without affecting the physical random
fluctuations.  An ad hoc solution for cold dark matter simulations may be to
de-noise them only over a range of scales that is adapted to the phase of
clustering: from the cell size to the size of the structures that have formed
latest (see Paper I).  Partial de-noising up to a given scale is implemented
in the code, but it is not tested in this context; the scale parameter is
$N_{\mathrm{t\,min}}$ (see Sect.\ 4.3).  An upper scale of $2^{n}$ cell sizes
corresponds to $N_{\mathrm{t\,min}}=N_{\mathrm{d}}/2^{n}$.  At the beginning
of the simulation, before the first virialized systems have formed, the upper
scale should match the smallest physical scale unaffected by the $N$-body
method, so $N_{\mathrm{t\,min}}$ should be set to $N_{\mathrm{d}}/2$ or
$N_{\mathrm{d}}/4$.  Analogous solutions may be found in the context of other
cosmological models.

\section{HOW TO USE THE CODE}

A flowchart of the code is illustrated in Fig.\ 14 (the symbols are standard;
see, e.g., Nyhoff \& Leestma 1997).  It summarizes the most useful
information given in Sects 4 and 5, without repeating the relevant
definitions.  In this section, we discuss such practical points in detail.
Supplementary information is given in the readme-file of the code
distribution.

   The code can be used for de-noising $N$-body simulations, and 1D--3D
standard data with Poissonian noise or additive white Gaussian noise.  It
contains include-files for many orthogonal and bi-orthogonal wavelet filters,
and also routines for the fast wavelet transform and its inverse.  The number
of data points should contain powers of two.  Data of different size can be
padded: add zeros, or extend the data so that their boundary values match
smoothly.  Smooth padding is better because it reduces wrap-around effects.
Data with multiplicative and/or coloured Gaussian noise can also be
de-noised.  In the first case, pre-process the data by taking their
logarithm, de-noise them in the usual way and post-process.  In the second
case, compute the standard deviation of noise on each scale from the
corresponding detail, and threshold the wavelet coefficients accordingly.

   We now explain how to include our add-on code in particle-mesh $N$-body
codes (e.g., Combes et al.\ 1990; Pfenniger \& Friedli 1993; Klypin \&
Holtzman 1997).  The proper de-noising subroutine should be called just after
the mass/charge assignment.  Note that the right argument is the number of
particles per cell in the active grid, not the mass/charge distribution in
the whole mesh.  Therefore the subroutine needs a simple interface for the
conversion of such arrays.  The specific form of interface depends on Fortran
details.  If there are various particle species, which represent components
with different collision properties, then each species can have its own type
of de-noising.  The case of polar grids is similar to the Cartesian case.  In
fact, for our purpose we can regard the space spanned by the cell indices as
Cartesian and the particle distribution defined there as evenly sampled.  In
addition, the boundary values of the particle distribution match smoothly,
except near the intersections between the radial boundaries and the
equatorial plane.  So smooth padding may be justified even if the number of
radial cells already contains a power of two.  In order to reduce wrap-around
effects significantly, the thickness of the padding layer should be
comparable to the size of the wavelet filters.  Note that such extra cells
are only used for de-noising purposes and do not enter into the $N$-body
computation itself.  The case of other grid geometries is analogous.  It is
not yet clear how to include our add-on code in other types of $N$-body
codes.

   Let us finally remark that the physical performance of the code depends on
how it is used.  In order to get a very good performance, follow the
guide-lines of Sects 4 and 5 and the practical advice of this section.  The
performance can be optimized in the case of galaxy simulations, since the
initial model is noisy and the theoretical particle distribution is known.
The degrees of freedom are the choice of the wavelet and the value of the
scale parameter (cf.\ Sects 4.2 and 4.3).  The optimization consists of a
simple trial: vary such degrees of freedom so as to get the largest
de-noising factor, and check the visual quality of the de-noised model.  In
the case of cosmological simulations of structure formation in the early
Universe, the value of the scale parameter is a critical issue (cf.\ Sect.\
5.3), while an appropriate choice of the wavelet may be guessed considering
the characteristics of such structures.  For example, in cold dark matter
simulations we would choose the `bior\,4.4' wavelet (cf.\ Fig.\ 1) since the
haloes that form are cuspy.  In the cases of plasma simulations and standard
data, we cannot give more specific instructions than those of Sects 4 and 5.

\section{CONCLUSIONS}

$N$-body simulations of structure formation in the early Universe, of
galaxies and plasmas are limited crucially by noise, whose effects are subtle
and not yet fully understood.  In Paper I, we have introduced an innovative
multi-scale method of noise reduction based on wavelets, which promises
marked advances in those contexts.  In this paper, we have discussed such a
method and its code implementation.  We have also explained how to include
our code in the simulator's $N$-body code, and how to use it for de-noising
standard data.  The code is available on request from the first author.  The
major conclusions of this paper are pointed out below.
\begin{enumerate}
   \item This is the first wavelet add-on code designed for de-noising
$N$-body simulations, and as such is meant to be a building block for more
elaborate de-noising codes.  We hope to have stimulated curiosity about the
uses of our code, and we challenge simulators to apply it to physical
problems where noise must be suppressed or controlled.
   \item The strength of the code is twofold.  It improves the performance of
simulations up to two orders of magnitude (cf.\ Paper I).  Besides, it allows
controlling the effects of noise: the $N$-body simulation can be made
equivalent to a simulation with a pre-assigned number of particles
$N_{\mathrm{pre}}$, for $N_{\mathrm{pre}}/N$ larger than unity and spanning
one order of magnitude.
   \item The weakness or rather small imperfection of the code is that
noise-generated instabilities are not reproduced very well, in contrast to
the induced dynamical evolution.  Obviously, errors may follow from an
incorrect use of the code.
   \item Finally, we believe that the performance of simulations can be
further improved with a more appropriate pre-processing of the data.
Fry\'zlewicz \& Nason (2004) have shown that the Haar-Fisz transformation is
better than the Anscombe transformation for pre-processing data with
Poissonian noise, and that the computational time is comparable (see also
Fryzlewicz 2003).  Work is in progress to investigate other relevant
properties and uses of this transformation, before including it in our code.
\end{enumerate}

\section*{ACKNOWLEDGMENTS}

The first author dedicates the code to his wife {\AA}sa and his children
Johan, Filip and Lukas; the code is named after them.  It is a great pleasure
to thank the referee, Alexander Knebe, for strong encouragement, valuable
suggestions and detailed comments.  We are very grateful to John Black,
Stefan Goedecker, Michael Joyce, Imre P\'{a}zsit, Daniel Pfenniger, Gustaf
Rydbeck, Peter Teuben and Hans van den Berg for strong support and useful
discussions.  We also acknowledge the financial support of the Swedish
Research Council and a grant by the `Solveig och Karl G Eliassons
Minnesfond'.  The paper was submitted with a delay of several months because
the first-author's computer was stolen twice and many important files were
lost.

\bsp

\label{lastpage}

\end{document}